\documentclass[aps,prl,preprint,superscriptaddress,nobibnotes]{revtex4}
\usepackage{amsmath,amssymb}

\begin{document}


\title{Family Unification, Exotic States and Light Magnetic Monopoles}


\author{Thomas W. Kephart}
\email[]{thomas.w.kephart@vanderbilt.edu}
\affiliation{Department of Physics and Astronomy,\\
Vanderbilt University, Nashville, TN 37325}
\author{ Chin-Aik  Lee }
\email[]{jlca@udel.edu}
 \author{Qaisar Shafi}
\email[]{shafi@bartol.udel.edu}

\affiliation{Bartol Research Institute, Department of Physics and 
Astronomy,\\
University of Delaware, Newark, DE 19716.}


\date{\today}

\begin{abstract}
Models with fermions in bifundamental representations can lead naturally to family unification as opposed to family replication.
Such models typically predict (exotic)
color singlet states with fractional electric charge, and
magnetic monopoles with multiple Dirac 
charge.
The  exotics may be at the TeV scale, and relatively light magnetic monopoles ($\gtrsim 10^7 GeV$) can be present in the galaxy with abundance near the Parker bound. We focus on three family $SU(4)\times SU(3)\times SU(3)$ models.
\end{abstract}


\maketitle


\section{Introduction}
Models with  fermions in bifundamental representations of product gauge groups of the general form 
$SU(a)\times SU(b)\times SU(c)$ have been studied for a variety of reasons over the last three decades
\cite{f1}.
The fermions must be free of gauge anomalies. The relevant gauge anomalies here are $SU(a)^3$, $SU(b)^3$ and $SU(c)^3$ (unless either a, b or c =2). So if the fermions are composed solely of bifundamental representations, they must be of the form
\begin{equation}\label{abc}
\frac{1}{d}[c(a,\bar{b},1)\oplus a(1,b,\bar{c})\oplus b(\bar{a},1,c)]
\end{equation}
or a multiple thereof. Here, $d$ is the lowest common divisor of $a,b$ and $c$. A simple example is the Trinification (TR) model \cite{trin} where $a=b=c=d=3$ and a single anomaly family lives in the bifundamental representations $R_{TR}=(3,\bar{3},1)\oplus (1,3,\bar{3})\oplus (\bar{3},1,3)$ of the gauge group $G_{TR}=SU(3)\times SU(3)\times SU(3)$.
Another example is the Pati-Salam (PS) model where the gauge group is $G_{PS}=SU(4)\times SU(2)\times SU(2)$ \cite{Pati:1974yy} and the fermions representations are $R_{PS}=(4,2,1)\oplus(\bar{4},1,2)$. This is an exception to the above general remarks since we do not need to include $(1, 2,\bar{2})$ because there are no $SU(2)^3$ gauge anomalies. This only happens when the gauge group contains $SU(2)$ subfactors. We still have $SU(2)$ global anomalies to worry about, but there are an even number of doublets, and so, the model is consistent.

In \cite{Kephart:2001ix} we introduced a model based on the gauge group $SU(4)\times SU(3)\times SU(3)$ (the 334-model). It is the minimal model  that contains both the 
(PS) model and the (TR) model. The 334-model contains features that are not present in many grand unified models. These include fractionally charged color singlet states and light magnetic monopoles. The richness of possible charge assignments were not explored in \cite{Kephart:2001ix}, but here we begin a more comprehensive analysis.

If string theory is to be the ultimate physical theory, then we must be able to extract standard model (SM) physics from it.
A number of attempts with this objective have been pursued, including Calabi-Yau compactifications of the heterotic string, which yield $E_{6}$ type $GUT$ theories, where holomorphic
deformations, Wilson loops, etc., can be used to reduce the gauge
symmetry. Orbifolding of  type $IIB$ strings on 
$AdS_{5}\times S^{5}$ can produce four dimensional conformal field theories ($CFT$s$)$
with gauge groups $\prod_{i} SU(Nd_{i})$ and bifundamental matter
\cite{Lawrence:1998ja}, \cite{Kachru:1998ys} and hence resemble the standard model. This is part of our motivation for studying the 334 model.

Here we again \cite{Kephart:2001ix} take a bottom-up approach and consider  models that are likely to be derivable
from orbifolded type $IIB$ strings, but our focus will be the resulting phenomenology, not on a string theory derivation of the model.
The models we study contains aspects of both Calabi-Yau and $AdS/CFT$
type string theory compactifications, and leads to a remarkably rich
set of phenomenologies.
It is well known that the (PS) model
and the 
(TR)
model are both contained in $E_{6}$ Grand Unification
\cite{Gursey:1976ki},
\cite{Achiman}. 
We will provide
additional coverings of PS and TR which do not embed in $E_{6}$, but
instead
{\it require} nontrivial family unification and are perhaps the minimal such example of a models
with
this property. We conclude this section with a brief review of PS and TR models. In the following sections, we present our models
and
then consider some of their phenomenological consequences.

The gauge group of the PS model is $G_{PS}=SU(4)\times SU(2)_{L}\times
SU(2)_{R}$.
Each fermion family lives in a set of bifundamental representation
\begin{equation}
(4,2,1)\oplus(\bar{4},1,2).
\end{equation}
This model embeds naturally in $SO(10)$ (as $G_{PS}/\mathbb{Z}_2$), where a fermion family combines into
a {\bf16} of $SO(10)$. Adding a $\mathbf{10} \oplus \mathbf{1}$ of fermions then allows unification into $E_{6}$,
where the fermions are now in a {\bf27}. The PS, $SO(10)$, and $E_6$ models are all chiral and anomaly free, family by family, and so a full
three-family \cite{f2}
model is gotten simply by replicating the first
family.

The TR model also has fermions in bifundamental representations 
\begin{equation}
(3,\bar{3},1)\oplus(\bar{3},1,3)\oplus(1,3,\bar{3})
\end{equation}
where now the gauge group is
$G_{TR}=SU(3)\times SU(3)\times SU(3).$ As $G_{TR}/\mathbb{Z}_3$ is a subgroup of $E_{6}$ and  (3) already
contains 27 fermionic states, the unification into $E_{6}$ is gotten simply by
adding the necessary gauge generators to extend $SU(3)^3$ to $E_{6}$. Again,
a single fermion family is anomaly free on its own, so we must add two more
families to agree with phenomenology.

\section{Review of the $SU(4)\times SU(3)\times SU(3)$ Model}

The smallest group that contains
both
the PS and TR models is not $E_{6}$ but $G=SU(4)\times SU(3)\times
SU(3)$,
which has 31 generators and has a rank of 7. Insisting on fermions in bifundamental
representations, we consider  $(4,\bar{3},1)$,
$(\bar{4}%
,1,3)$, and $(1,3,\bar{3})$. We {cannot} take one of each to form a
family,
since this would be anomalous. The minimal anomaly free choice is \cite{Kephart:2001ix}

\begin{equation}
3(4,\bar{3},1)\oplus 3(\bar{4},1,3)\oplus 4(1,3,\bar{3}).
\end{equation}
If we break the $SU(4)$ to $SU(3)$, then (4) becomes

\begin{equation}
3[(3,\bar{3},1)\oplus(\bar{3},1,3)\oplus(1,3,\bar{3})]\oplus(1,\bar{3},1)\oplus(1,1,3)\oplus(1,3,\bar{3})
\end{equation}
which contains three TR families plus a few additional particles. Hence the
simplest set of anomaly free chiral bilinear representation $[i.e.,(4)]$
contains
three families. This is a true family unification, instead of a model where the 
second and third families are gotten from merely
replicating the first. Examples of the latter include minimal $SU(5)$, $SO(10)$ with fermionic {\bf 16}s, and  $E_6$ grand unification with fermions in {\bf 27}s, in addition to the PS and TR models.

The full analysis of the  334-model \cite{footnote}
requires the study of all the various patterns of spontaneous symmetry breaking (SSB), and the 
charge assignments these lead
to,
plus the phenomenological implication of the ``extra" fermions. Typically there exist
fractional charged color singlets
\cite{Kim:1980yk},  
\cite{Goldberg:1981jt} in these models, and 
hence the minimal
monopole change will be the inverse of this minimal fraction
times the Dirac charge.

As we shall see, there are only three inequivalent possibilities for embedding color and weak isospin of the standard model in $%
SU(4)\times SU(3)\times SU(3)$ (if we ignore the diagonal subgroups). 
The embedding of weak hypercharge is more complicated. Consider the
breaking $%
SU(4)\times SU(3)_{A}\times SU(3)_{B}\rightarrow SU(4)\times
\frac{SU(2)_{L}\times U(1)_{A}}{\mathbb{Z}_2}\times \frac{SU(2)_{R}\times U(1)_{B}}{\mathbb{Z}_2}$. If we then
break $U(1)_{A}$ and $U(1)_{B}$ completely, the hypercharge must be $%
Y=T_{3R}+(B-L)$, where $T_{3R}$ is the diagonal generator of
$SU(2)_{R}$,
and $B-L$ generates the $U(1)$ that is in $SU(4)$ but not in
$SU(3)_{C}$.
However, there are many other possibilities for the embedding of $U(1)_{Y}$.
These are similar to the well-known
flipped
models
\cite{DeRujula:1980qc},
\cite{Kyae:0510105}.
One obvious choice is to break $SU(4)$ to $SU(3)_{C}$ and then
$Y$
could be the trinification choice from $SU(3)_{L}\times SU(3)_{R}$.
Trinification has a standard hypercharge assignment, but
this could be flipped. Also moving
$SU(2)_{W}$ from $SU(4)$ to an $SU(3)$ of the 334-model corresponds to an isoflipped
model
\cite{Kephart:1989az}. (There are even more choices where family members move around but the charges of the extra 
fermions change. These``transflipped" models will be described
in the next section.)
Here we restrict ourselves to the standard hypercharge embeddings that generate PS and TR models, but keep in mind that flipping may offer other opportunities.
(Note, embedding  $SU(3)_{C}$ and/or
$SU(2)_{W}$ in some diagonal subgroup within the 334-model, and this includes the ``diagonal embedding'' 
$SU(2)_{W}=SU(2)_{diag} \subset SU(2)\times SU(2) \subset SU(4)$,
leads to vectorlike fermions, and this route is incompatible with SM phenomenology.)

We now begin our review of the most straightforward embedding followed in \cite{Kephart:2001ix}
 which leads to the TR and PS models. In the next section we give a general analysis of all embeddings where the patterns of spontaneous symmetry breaking can be consistent with SM phenomenology.

The standard PS version of the 334-model, has fermions
\begin{equation}
3[(4,\bar{2},1)\oplus(4,\bar{1},1)]\oplus 3[(\bar{4},1,2)\oplus(\bar{4},1,1)]\oplus 4[(1,2,\bar{2})
\oplus (1,2,1)\oplus (1,1,\bar{2})\oplus (1,1,1)]
\end{equation}
Only the three PS families remain chiral, while the
extra (exotic) vectorlike fermions obtain masses from Higgs VEVs at the G breaking scale. This will be discussed in more detail in the section on the Higgs sector.  
Breaking $G_{PS}$
to the standard model gauge group $G_{SM}=\frac{SU(3)_{C}\times SU(2)_{W}\times U(1)_{Y}}{\mathbb{Z}_3}$,
we
find
\begin{eqnarray*}
&&3[(3,2)_{\frac{1}{6}}+(1,2)_{-\frac{1}{2}}+(3,1)_{\frac{1}{6}}+(1,1)_{-\frac{1}{2}}] \\
&&+3[(\bar{3},1)_{\frac{1}{3}}+(\bar{3},1)_{-\frac{2}{3}%
}+(1,1)_{1}+(1,1)_{0}+(\bar{3},1)_{-\frac{1}{6}}+(1,1)_{\frac{1}{2}}] \\
&&+4[(1,2)_{-\frac{1}{2}}+(1,2)_{\frac{1}{2}}+(1,2)_{0}+(1,1)_{-\frac{1}{2}}
+(1,1)_{\frac{1}{2}}+(1,1)_{0}.]
\end{eqnarray*}
 As expected, we are left with three standard model families, plus three
right-handed neutrinos, needed to form the three PS families, plus the following extra fermion states:
\begin{eqnarray*}
&&3[(3,1)_{\frac{1}{6}}+(\bar{3},1)_{-\frac{1}{6}}]+4[(1,2)_{\frac{1}{2}}
+(1,2)_{-\frac{1}{2}}]+7[(1,1)_{\frac{1}{2}}+(1,1)_{-\frac{1}{2}}] \\
&&+4(1,2)_{0}+4(1,1)_{0}.
\end{eqnarray*}
\newline
Electric charge is now quantized in units of
$\frac{1}{2}$,
so the magnetic monopole must have a minimum
charge of two from the Dirac quantization condition.

Care must be taken with the direct route to the TR model. This is due to a subtlety that arises in breaking $SU(4)$ to $SU(3)$.
With the standard trinification charge assignments, one
finds that  massless charged quarks and leptons appear in the spectrum-- a conflict with phenomenology. To avoid this we must  include additional fermion multiplets at the
334 level. Let us see how this works. Two inequivalent cases must be
considered:
(i) $SU(3)_{C}$ embedded in $SU(4)$, or (ii) $SU(3)_{C}$ identified with
an $%
SU(3)$ of the 334-model. In both cases at the trinification level we
begin with
fermions as in (4).

For case (i), we have the three standard families
plus
\begin{equation}
R_{E}=3(1,\bar{3},1)\oplus 3(1,1,3)\oplus(1,3,\bar{3})
\end{equation}
under $SU(3)_{C}\times SU(3)_{L}\times SU(3)_{R}$. Hence all the extra
states are leptonic. Then for $SU(3)_{L}\times SU(3)_{R}\rightarrow  
\frac{SU(2)_{L}\times U(1)_{L}}{\mathbb{Z}_2}\times U(1)_{R}$ where we identify $U(1)_{R}$
with
the diagonal generator $Y_{R}=diag(1,1,-2)$ of $SU(3)_{R}$ and likewise
$%
U(1)_{L}$ with the  generator $Y_{L}=diag(1,1,-2)$ of $SU(3)_{L}$, we can choose the hypercharge to be
$Y=\frac{1}{6}Y_{L}+\frac{1}{3%
}Y_{R}$. The families just have the standard ${\bf 27}$ of $E_{6}$
charges,
while the new leptons are
\begin{equation}
5(1,2)_{-\frac{1}{6}}+(1,2)_{\frac{5}{6}}+10(1,1)_{\frac{1}{3}%
}+5(1,1)_{-\frac{2}{3}}
\end{equation}
These states are still chiral, and the only way to give them mass would
be with a VEV from an electrically charged Higgs. As this must obviously be
avoided, an alternative, if we relax our restriction on only having bifundamental fermions, 
is to arrange these particles to be vectorlike
by adding the conjugate, but anomaly free chiral multiplets
\begin{equation}
\bar{R}_{E}=3(1,3,1)+3(1,1,\bar{3})+(1,\bar{3},3)
\end{equation}
at the 334 level. But this particular combination of $R_{E}$ and $\bar{R}_{E}$ contains the vectorlike pair 
of bifundamentals $(1,3,\bar{3})\oplus (1,\bar{3},3)$
which is also anomaly free since it is vectorlike. Barring any additional 
symmetry or some other mechanism, we would expect this pair to acquire 
a mass much higher than the 334 breaking scale, which will effectively 
lead to its decoupling. This is possible because there is a smaller anomaly 
free combination (since we now have fundamentals in addition to bifundamentals).
\begin{equation}
3(4,\bar{3},1)\oplus 3(\bar{4},1,3)\oplus 3(1,3,\bar{3})\oplus 3(1,3,1)\oplus 3(1,1,\bar{3})\label{noFanU}
\end{equation}
But this is none other than three copies of the anomaly free combination:
\begin{equation}
(4,\bar{3},1)\oplus (\bar{4},1,3)\oplus (1,3,\bar{3})\oplus (1,3,1)\oplus (1,1,\bar{3})
\end{equation}
 Unfortunately, this means that the case where 334 breaks down via TR does not explain why there are three families.

Now, with fermions as in (\ref{noFanU}), upon breaking $G\rightarrow G_{TR}$ at some high scale $\langle\phi\rangle$, 
the chiral families stay massless while 
the extra fermions acquire mass terms  
of the form $h\langle\phi\rangle$ $\bar{R}_{E}R_{E}$, where $h$ is a typical Yukawa coupling constant.
Hence, the fractionally charged leptons
become heavy compared to the family fermions. Let us summarize the
extra vectorlike leptons. There are five doublets with electric charges
$\pm
\frac{1}{3}$ and $\mp \frac{2}{3}$ , two doublets with electric charge
$\pm
\frac{1}{3}$ and $\pm \frac{4}{3}$, ten singlets with $\pm \frac{1}{3}$
charges, and five singlets with  $\mp \frac{2}{3}$ charges. The minimal
monopole charge is three.

\bigskip

For case (ii), some of the extra states will be colored. In terms of
$SU_{C}(3)%
\times SU_{L}(3)\times SU_{R}(3)$, they are
\begin{equation}
S_{E}=3(3,1,1)+3(1,1,\bar{3})+(\bar{3},1,3).
\end{equation}
The hypercharge in unchanged 
from case (1) (it is still $Y=\frac{1}{6}Y_{L}+\frac{1}{3%
}Y_{R}$), so now we find:
\begin{equation}
S_{E}=3(3,1)_0+3(1,2)_{-{\frac{1}{3}}}+3(1,1)_{\frac{2}{3}}+(\bar{3},2)_{\frac{1}{3}}%
+(\bar{3},1)_{-{\frac{2}{3}}}.
\end{equation}
Again we must add conjugate states
\begin{equation}
\bar{S}_{E}=3(\bar{3},1,1)+3(1,1,3)+(3,1,\bar{3}),
\end{equation}
and this allows masses for the exotics at a scale higher than the family masses.

With this review under our belt we are now ready to survey the complete list 
of allowed patterns of spontaneous symmetry breaking for the 334 model that 
lead to standard model physics plus extended families of fermions. As there 
are alternative hypercharge and weak isospin assignments (the flipped and 
isoflipped models) in $SO(10)$ and $E_6$, so there are alternative embeddings 
of hypercharge and weak isospin  in $SU(4)\times SU(3)_L\times SU(3)_R$ that 
are not the same as the PS or TR assignments discussed above. Recall that we 
will call these alternative hypercharge and weak isospin assignments ``transflipped 
334 models." In this case the flipping rearranges the location of the family members 
while keeping their charges fixed, but the extra fermions change 
both their location and charge. This is more general than the behavior of flipping 
in $SO(10)$ or $E_6$ where rearrangement takes place, but charges of extra fermions 
remain fixed.

\section{Classification of inequivalent 334-models}

There are three inequivalent ways to embed $SU(3)_C$ and $SU(2)_W$ in $SU(4)\times SU(3)_L\times SU(3)_R$ that can 
lead to the correct fermion families in the standard model group $SU(3)_C\times SU(2)_W\times U(1)_Y$.\\

\begin{enumerate}
\item[(i)] Embed $SU(3)_C$ in $SU(4)$ and $SU(2)_W$ in either 
$SU(3)_L$ or $SU(3)_R$.
\item[(ii)] Identify $SU(3)_C$ with 
$SU(3)_L$ and embed $SU(2)_W$ in $SU(3)_R$ (or vice versa).
\item[(iii)] Identify $SU(3)_C$ with either 
$SU(3)_L$ or $SU(3)_R$ and let $SU(2)_W\subset SU(4)$.
\end{enumerate}

In all cases the fermion representations at the 334 unification scale are\\ $3(4,\bar{3},1)\oplus
3(\bar{4},1,3)\oplus 4(1,3,\bar{3})$, but in some circumstances it may be necessary to include additional fundamental or bifundamental representations to avoid unwanted massless charged particles.

In case (i) the initial $SU(4)$ contains a diagonal 
generator $\Lambda_{15}=\frac{1}{\sqrt{6}}diag(1,1,1,-3)$ orthogonal to $SU(3)_C$
\cite{f3}.
  With $SU(2)_W$ in $SU(3)_L$ we have another commuting generator 
$\lambda_{8}^L=\frac{1}{\sqrt{3}}diag(1,1,-2)$, and there are two more diagonal $U(1)$ generators 
$\lambda_{3}^L=diag(1,-1,0)$ and $\lambda_{8}^R=\frac{1}{\sqrt{3}}diag(1,1,-2)$ in $SU(3)_R$. We now require the weak hypercharge $U(1)_Y$ be generated by a linear combination of $\Lambda_{15}$, $\lambda_{8}^L$, $\lambda_{3}^R$ and 
$\lambda_{8}^R$ such that we arrive at three families plus additional states. We can proceed systematically by following the decomposition of the fermion bifundamentals for 
$$SU(4)\times SU(3)_L\times SU(3)_R \rightarrow SU(3)_C\times U(1)_{\Lambda_{15}}\times SU(2)_W \times U(1)_{\lambda_{8}^L} \times U(1)_{\lambda_{3}^R} \times U(1)_{\lambda_{8}^R}$$ to find
\begin{equation}
(4,\bar{3},1) \rightarrow (3,\bar{2},1)_{1-100}+(1,\bar{2},1)_{-3-100}+(3,1,1)_{1200}+(1,1,1)_{-3200}
\end{equation}
\begin{eqnarray}
(1,3,\bar{3}) \rightarrow   (1,2,1)_{01-1-1}+ (1,2,1)_{011-1}+(1,2,1)_{0102}\nonumber\\
~~~~+(1,1,1)_{0-2-1-1}+(1,1,1)_{0-21-1}+(1,1,1)_{0-202}
\end{eqnarray}
\begin{eqnarray}
(\bar{4},1,3) \rightarrow    (\bar{3},1,1)_{-1011}+(\bar{3},1,1)_{-10-11}+(\bar{3},1,1)_{-100-2}\nonumber\\
~~~~+(1,1,1)_{3011}+(1,1,1)_{30-11}+(1,1,1)_{300-2}
\end{eqnarray}
 
We now must solve for the generator of $U(1)_Y$. We do this by solving for the coefficients in
\begin{equation}
Y=a\Lambda_{15}+b\lambda_{8}^L+c\lambda_{3}^R+d\lambda_{8}^R
\end{equation}
by identifying the families in the 334 bifundamentals. To start the process, note
that in a standard family there is only one quark doublet, i.e., the $(3,2)_{\frac{1}{6}}$. Therefore, comparing with the decomposition of the $(4,\bar{3},1)$ we must have 
\begin{equation}
a-b=1/6
\end{equation}
where we take $a,b,c$ and $d$ to be the coefficients of the unnormalized version of the $U(1)$ generators  $\Lambda_{15}$, $\lambda_{8}^L$, $\lambda_{3}^R$ and 
$\lambda_{8}^R$.

There are three sets of $(\bar3,1)$s, so we can get two more conditions on the coefficients in the definition of $Y$ by setting one of the three corresponding linear combinations (i.e., $-a+c+d$, $-a-c+d$, or $-a-2d$) equal to the family hypercharge for the $(\bar3,1)_{-2/3}$ and another to the hypercharge of the $(\bar3,1)_{1/3}$.
At this stage there is still one free parameter, but we need to check that sufficient freedom remains to have three $(1,2)_{-1/2}$s and three $(1,1)_1$s.
 By setting lepton doublet charges to 
${-\frac{1}{2}}$ or singlet charges to $1$ we find the allowed solutions summarized in Table I.

\bigskip

\begin{center}
\begin{tabular}{|l|l|l|l|l|}
\hline
a & b & c & d & type \\ \hline\hline
0 & $-1/6$ & -1/2 & -1/6 & TR \\ \hline
1/6 & 0 & 1/2 & 0 & PS \\ \hline
-4/3 & -3/2 & -1/2 & -3/2 & I \\ \hline
-1/3 & -1/2 & 1 & 0 & II \\ \hline
5/12 & 1/4 & -1/2 & 1/4 & III \\ \hline 
2/3 & 1/2 & 1 & 0 & IV \\ \hline
\end{tabular}
\end{center}
\bigskip
Table I: The six possible models of embedding class (i). Both the Pati-Salam model and the trinification model are of this class.

\bigskip

Other choices of coefficients correspond to trivial flippings. For instance, there are three equivalent choices for 
the overall diagonal generator of  $SU(3)_R$ that enters $Y$, and they are given by (1.) $c=-1/2$ and $d=-1/6$, (2.) $c=1/2$ and $d=-1/6$, and (3.) $d=1/3$ with $c=0$. (In $E_6$  this type of flipping is nontrivial if the spontaneous symmetry breaking is such that the states end up in different irreps of the decomposition, e.g., different routes through $SU(5)$ irreps.) As we mentioned above, the 334 model has nontrivial transflipping where the family states move about while the charges of the extra states change, and these are model I through IV in Table I. As we already know the content of the TR and PS models we derived, let us summarize the fermion contents of the new models I through IV. They are

Case I:  

The coefficients are 

\begin{equation}
(a,b,c,d)=(-4/3,-3/2,-1/2,-3/2).
\end{equation}

They yield a weak hypercharge operator

\begin{equation}
Y_I=\underbrace{\left(\begin{array}{cccc}-\frac{4}{3}&0&0&0\\0&-\frac{4}{3}&0&0\\0&0&-\frac{4}{3}&0\\0&0&0&4\end{array}\right)}_{SU(4)}+\underbrace{\left(\begin{array}{ccc}-\frac{3}{2}&0&0\\0&-\frac{3}{2}&0\\0&0&3\end{array}\right)}_{SU(3)_L}+\underbrace{\left(\begin{array}{ccc}-2&0&0\\0&-1&0\\0&0&3\end{array}\right)}_{SU(3)_R},
\end{equation}

and fermions
 
 \begin{equation}
(4,\bar{3},1) \rightarrow \underline{(3,{2})_{1/6}}+(1,{2})_{11/2}+(3,1)_{-13/3}+\underline{(1,1)_{1}}
\end{equation}
\begin{eqnarray}
(1,3,\bar{3}) \rightarrow   (1,2)_{1/2}+ \underline{(1,2)_{-1/2}}+(1,2)_{-9/2}\nonumber\\
~~~~+(1,1)_{5}+(1,1)_{4}+(1,1)_{0}
\end{eqnarray}
\begin{eqnarray}
(\bar{4},1,3) \rightarrow    \underline{(\bar{3},1)_{-2/3}}+\underline{(\bar{3},1)_{1/3}}+(\bar{3},1)_{13/3}\nonumber\\
~~~~+(1,1)_{-6}+(1,1)_{-5}+(1,1)_{-1},
\end{eqnarray}
 where we have underlined the locations of members of the standard families.
 
Case II:

Now the coefficients are

\begin{equation}
(a,b,c,d)=(-1/3,-1/2,1,0),
\end{equation}

 yielding 
 
\begin{equation}
Y_{II}=\underbrace{\left(\begin{array}{cccc}-\frac{1}{3}&0&0&0\\0&-\frac{1}{3}&0&0\\0&0&-\frac{1}{3}&0\\0&0&0&1\end{array}\right)}_{SU(4)}+\underbrace{\left(\begin{array}{ccc}-\frac{1}{2}&0&0\\0&-\frac{1}{2}&0\\0&0&1\end{array}\right)}_{SU(3)_L}+\underbrace{\left(\begin{array}{ccc}-1&0&0\\0&0&0\\0&0&1\end{array}\right)}_{SU(3)_R},
\end{equation}

and fermions
 
 \begin{equation}
(4,\bar{3},1) \rightarrow \underline{(3,{2})_{1/6}}+(1,{2})_{3/2}+(3,1)_{-4/3}+(1,1)_{0}
\end{equation}
\begin{eqnarray}
(1,3,\bar{3}) \rightarrow   (1,2)_{-3/2}+ (1,2)_{1/2}+\underline{(1,2)_{-1/2}}\nonumber\\
~~~~+(1,1)_{0}+(1,1)_{0}+\underline{(1,1)_{1}}
\end{eqnarray}
\begin{eqnarray}
(\bar{4},1,3) \rightarrow   (\bar{3},1)_{4/3}+\underline{(\bar{3},1)_{-2/3}}+\underline{(\bar{3},1)_{1/3}}\nonumber\\ 
~~~~+(1,1)_{0}+(1,1)_{-2}+(1,1)_{-1},
\end{eqnarray}
 where we have again underlined the locations of members of the standard families. Note some of them have changed location from model I and the extra fermions have changed their locations as well as their charges.

Case III:

The coefficients are

\begin{equation}
(a,b,c,d)=(5/12,1/4,-1/2,1/4),
\end{equation}

leading to the hypercharge operator
 
\begin{equation}
Y_{III}=\underbrace{\left(\begin{array}{cccc}\frac{5}{12}&0&0&0\\0&\frac{5}{12}&0&0\\0&0&\frac{5}{12}&0\\0&0&0&-\frac{5}{4}\end{array}\right)}_{SU(4)}+\underbrace{\left(\begin{array}{ccc}\frac{1}{4}&0&0\\0&\frac{1}{4}&0\\0&0&-\frac{1}{2}\end{array}\right)}_{SU(3)_L}+\underbrace{\left(\begin{array}{ccc}-\frac{1}{4}&0&0\\0&\frac{3}{4}&0\\0&0&-\frac{1}{2}\end{array}\right),}_{SU(3)_R}
\end{equation}
 
and fermions

 \begin{equation}
(4,\bar{3},1) \rightarrow \underline{(3,{2})_{1/6}}+(1,{2})_{-3/2}+(3,1)_{11/12}+(1,1)_{-3/4}
\end{equation}
\begin{eqnarray}
(1,3,\bar{3}) \rightarrow   (1,2)_{1/2}+ \underline{(1,2)_{-1/2}}+(1,2)_{3/4}\nonumber\\
~~~~+(1,1)_{-1/4}+(1,1)_{-5/4}+(1,1)_{0}
\end{eqnarray}
\begin{eqnarray}
(\bar{4},1,3) \rightarrow    \underline{(\bar{3},1)_{-2/3}}+\underline{(\bar{3},1)_{1/3}}+(\bar{3},1)_{-11/12}\nonumber\\
~~~~+\underline{(1,1)_{1}}+(1,1)_{2}+(1,1)_{3/4},
\end{eqnarray}
 with  the locations of standard family members underlined. Remarkably, except for the positron, they are in the same places as they were in model II. The charges of the extra fermions have changes as expected.

Case IV:

This time, the coefficients 

\begin{equation}
(a,b,c,d)=(2/3,1/2,0,1)
\end{equation}

yield
 
\begin{equation}
Y_{IV}=\underbrace{\left(\begin{array}{cccc}\frac{2}{3}&0&0&0\\0&\frac{2}{3}&0&0\\0&0&\frac{2}{3}&0\\0&0&0&-2\end{array}\right)}_{SU(4)}+\underbrace{\left(\begin{array}{ccc}\frac{1}{2}&0&0\\0&\frac{1}{2}&0\\0&0&-1\end{array}\right)}_{SU(3)_L}+\underbrace{\left(\begin{array}{ccc}0&0&0\\0&1&0\\0&0&-1\end{array}\right),}_{SU(3)_R}
\end{equation}
 
and fermions

 \begin{equation}
(4,\bar{3},1) \rightarrow \underline{(3,{2})_{1/6}}+(1,{2})_{-/2}+(3,1)_{5/3}+(1,1)_{-1},
\end{equation}
\begin{eqnarray}
(1,3,\bar{3}) \rightarrow   \underline{(1,2)_{-1/2}}+ (1,2)_{1/2}+(1,2)_{1/2}\nonumber\\
~~~~+(1,1)_{-2}+(1,1)_{0}+(1,1)_{-1},
\end{eqnarray}
\begin{eqnarray}
(\bar{4},1,3) \rightarrow    \underline{(\bar{3},1)_{1/3}}+(\bar{3},1)_{5/3}+\underline{(\bar{3},1)_{-2/3}}\nonumber\\
~~~~+(1,1)_{3}+\underline{(1,1)_{1}}+(1,1)_{2},
\end{eqnarray}
 where the locations of members of the standard families have moved and the extra charges have changed values once again. In all cases the standard model family is distributed amongst all three types of initial 334 model bifundamental fermion representations.

As we have seen, embedding $SU(3)_C$ in $SU(4)$ (embedding class (i)) can lead to the PS model, but this is certainly not the case for embedding classes (ii) and (iii). However, the TR model is allowed by all three embedding classes. We now move on the Class (ii) where $SU(3)_C$ is in an $SU(3)$ of the 334 model and $SU(2)_W$ is in the other $SU(3)$. The fermions decompose as
 
  \begin{equation}
(4,\bar{3},1) \rightarrow  (\bar{3},1)_{-1-1-10}+(\bar{3},1)_{1-1-10}+ (\bar{3},1)_{02-10}+(\bar{3},1)_{0030},
\end{equation}
  \begin{equation}
(1,3,\bar{3}) \rightarrow  (3,{2})_{000-1}+ +(3,1)_{000-2},
\end{equation}
 \begin{eqnarray}
(\bar{4},1,3) \rightarrow    (1,2)_{1111}+(1,1)_{111-2}\nonumber\\
~~~~~~+(1,2)_{-1111}+(1,1)_{-111-2}\nonumber\\
~~~~~~+(1,2)_{0-211}+(1,1)_{0-21-2}\nonumber\\
~~~~~~+(1,2)_{00-31}+(1,1)_{00-3-2},
\end{eqnarray}
and the generator of $U(1)_Y$ is
\begin{equation}
Y={\tilde a}\Lambda_{3}+{\tilde b}\Lambda_{8}^L+{\tilde c}\Lambda_{15}^R+{\tilde d}\lambda_{8}^R.
\end{equation}
We start the process of identifying the families in the 334 bifundamentals by again noting
there is only one type of quark doublet. Therefore, comparing with the decomposition of the $(4,\bar{3},1)$ we must have ${\tilde d}=-1/6$. Again requiring the existence of 
$(\bar{3},1)$s with hypercharges $1/3$ and $-2/3$ and leptons $(1,2)_{-1/2}$ and $(1,1)_1$
leads to a set of equations for the coefficients in (43) with solutions summarized in Table II.

\bigskip

\begin{center}
\begin{tabular}{|l|l|l|l|l|}
\hline
${\tilde a}$ & ${\tilde b}$ & ${\tilde c}$ & ${\tilde d}$ & type \\ \hline\hline
0 & $1/3$ & 0 & -1/6 & TR \\ \hline
1/2 & $\frac{1-2x}{6}$ & $\frac{x}{3}$ & -1/6 & X \\ \hline
\end{tabular}

\bigskip

Table II: Class (ii) models.
\end{center}
\bigskip

As expected the TR model is a solution, however,  the ``X" model is a surprise. We find we can satisfy all the conditions necessary to fix the charge of all standard model family members without the need to specify the value of the parameter $x$ in the hypercharge,
\begin{equation}
Y=\underbrace{\left(\begin{array}{cccc}\frac{2}{3}&0&0&0\\0&-\frac{1}{3}&0&0\\0&0&x-\frac{1}{3}&0\\0&0&0&-x\end{array}\right)}_{SU(4)}+\underbrace{\left(\begin{array}{ccc}-\frac{1}{6}&0&0\\0&-\frac{1}{6}&0\\0&0&\frac{1}{3}\end{array}\right)}_{SU(3)_R}.
\end{equation}
I.e., the generator 
$X = diag(0,0,1,-1)$ of $U(1)_X\subset SU(4)$ is still at our disposal. Before discussing this model, let us display the fermions:
 \begin{equation}
(4,\bar{3},1) \rightarrow  \underline{(\bar{3},1)_{-2/3}}+\underline{(\bar{3},1)_{1/3}}+ (\bar{3},1)_{1/3-x}+(\bar{3},1)_{x},
\end{equation}
  \begin{equation}
(1,3,\bar{3}) \rightarrow  \underline{(3,{2})_{1/6}}+(3,1)_{-1/3},
\end{equation}
 \begin{eqnarray}
(\bar{4},1,3) \rightarrow    (1,2)_{1/2}+\underline{(1,1)_{1}}\nonumber\\
~~~~~~+\underline{(1,2)_{-1/2}}+\underline{(1,1)_{0}}\nonumber\\
~~~~~~+(1,2)_{x-1/2}+(1,1)_{x}\nonumber\\
~~~~~~+(1,2)_{-x-1/6}+(1,1)_{-x+1/3}.
\end{eqnarray}
We find the three complete standard model families (underlined states) without specifying $x$. There are several choices for $x$ that can be used to generate a flipped model.  $x=1, 0,-2/3 $ or $1/3$ flip the $(\bar{3},1)$s,
$x=0$ flips the $(1,2)$s, and $x=3/2,1,4,3$ or $-7/6$ flips the $(1,1)$s. The model does not, a priori, require complete charge quantization of the standard families relative to the extra fermions.
In any model where $U(1)_X$ is spontaneously broken by  vacuum expectation values for Higgs scalars in a representation of $SU(4)$, electric charge will end up quantized depending on the charges of that  representation. However, it is not necessary for $U(1)_X$ to be spontaneously broken since all of the standard model fields are neutral under it, turning $U(1)_X$ into a hidden sector gauge symmetry.

To see this, define the generator $Z$ by

\begin{equation}
Z=\underbrace{\begin{pmatrix}\frac{2}{3}&0&0&0\\0&-\frac{1}{3}&0&0\\0&0&-\frac{1}{3}&0\\0&0&0&0\end{pmatrix}}_{SU(4)}+\underbrace{\begin{pmatrix}-\frac{1}{6}&0&0\\0&-\frac{1}{6}&0\\0&0&\frac{1}{3}\end{pmatrix}}_{SU(3)_R}
\end{equation}
Then, $\frac{Y}{2}=Z+xX$, which means that the subgroup generated by Y is a subgroup of $U(1)_Z \times U(1)_X$ and that subgroup cannot possibly be $U(1)_X$. If $x$ is a rational number, then that subgroup is isomorphic to $U(1)$ and we can call it $U(1)_Y$ and the hypercharges, and hence the electric charges are quantized. On the other hand, if $x$ is irrational, then that subgroup is dense in $U(1)_Z \times U(1)_X$. Since the model that we are dealing with is continuous under symmetry transformations, if we have an infinite sequence of gauge transformations leaving the vacuum
\cite{fx}
invariant that converges to some gauge transformation, then that gauge transformation also leaves the vacuum invariant. So, if $Y$ is unbroken up to the electroweak breaking scale, then so is the closure of the subgroup generated by it; $U(1)_Z \times U(1)_X$ is also unbroken right up to the electroweak scale. The electric charge operator is $Q=\frac{Y}{2}+\frac{I}{2}$, where $I$ is the weak isospin operator. Let us define $V=Z+\frac{I}{2}$, such that $Q=V+xX$. Since the electromagnetic symmetry is unbroken, by the same reasoning, this also means that $U(1)_V \times U(1)_X$ is also unbroken.
Nothing in this discussion depends upon the details of the symmetry breaking (Higgs or dynamical).

However, there will be cross-coupling terms for the gauge kinetic terms. Let $A_Z$ be the gauge field for $U(1)_{Z}$ and $A_X$ be the gauge field for $U(1)_{X}$, with $F_Z$ and $F_X$ their respective field stregnths. The gauge kinetic term will have the generic form

\begin{equation}
-\frac{1}{4 g_Z^2} F_Z^{\mu\nu}F_{Z \mu \nu}-\frac{1}{4 g_Y^2} F_X^{\mu \nu} F_{X \mu \nu} - \frac{\alpha}{2} F_Z^{\mu \nu} F_{X \mu \nu}
\end{equation}
where $\alpha$ is some dimensionless coefficient that receives contributions from radiative corrections.

By taking linear combinations $X$ and $Y'=Z+\beta X$, we can diagonalize the gauge kinetic terms with a proper choice of  $\beta$, but in that case, the $Y'$ charges of SM fields and the exotic fields will be incommensurable. Since all the SM fields are neutral under $X$, $A_X$ decouples from the low energy physics and so, it is not a problem that it remains unbroken. However, it is essential that all exotics with a nonzero $X$ charge have large mass, since they couple to both the standard model gauge fields and $A_X$. This probably means that our current model does not work phenomenologically because, so far we have not been able to find a way to make all the exotic fermions vectorlike. However, this mechanism works generically for any GUT theory which contains $U(1)_Z \times U(1)_X$ with all the standard model fields being $X$-neutral and the $Z$ charges of all the standard model fields coinciding with their hypercharges.

We are now ready to continue on to class (iii) models where $SU(2)$ is embedded in $SU(4)$. We can write the hypercharge operator as
\begin{equation}
Y ={\hat a}\Lambda_{8'}+{\hat b}\Lambda_{15'}+{\hat c}\lambda_{3}^R+{\hat d}\lambda_{8}^R,
\end{equation}
where we define $\Lambda_{8'}=diag(1,1,-1,-1)$ and $\Lambda_{15'}=diag(0,0,1,-1)$.
The fermions now decompose as (writing the representations in the order ($SU(3)_C,SU(2)_W$) to agree with the previous notation)
 
  \begin{equation}
(\bar{4},1,3) \rightarrow  (3,{2})_{-1000}+ (3,1)_{1-100}+ (3,1)_{1100}.
\end{equation}
  \begin{equation}
(1,3,\bar{3}) \rightarrow  (\bar{3},1)_{0011}+(\bar{3},1)_{00-11}+ (\bar{3},1)_{000-2}.
\end{equation}
 \begin{eqnarray}
(4,\bar{3},1) \rightarrow    (1,2)_{10-1-1}+(1,2)_{101-1}+(1,2)_{1002}\nonumber\\
~~~~~~+(1,1)_{-11-1-1}+(1,1)_{-111-1}+(1,1)_{-110-2}\nonumber\\
~~~~~~+(1,1)_{-1-1-1-1}+(1,1)_{-1-11-1}+(1,1)_{-1-102}.
\end{eqnarray}

The usual process of requiring the  
 quark doublets to have hypercharge ${\frac{1}{6}}$,  the   
$(\bar{3},1)$s have hypercharges $1/3$ and $-2/3$ and leptons $(1,2)_{-1/2}$ and $(1,1)_1$
leads to two inequivalent models where the hypercharge operator is either
 
 \begin{equation}
Y_{TR}=\underbrace{\left(\begin{array}{cccc}-\frac{1}{6}&0&0&0\\0&-\frac{1}{6}&0&0\\0&0&\frac{1}{3}&0\\0&0&0&0\end{array}\right)}_{SU(4)}+\underbrace{\left(\begin{array}{ccc}-\frac{2}{3}&0&0\\0&\frac{1}{3}&0\\0&0&\frac{1}{3}\end{array}\right)}_{SU(3)_R}
\end{equation}
 
 or 
 
\begin{equation}
Y_{V}=\underbrace{\left(\begin{array}{cccc}-\frac{1}{6}&0&0&0\\0&-\frac{1}{6}&0&0\\0&0&\frac{4}{3}&0\\0&0&0&-1\end{array}\right)}_{SU(4)}+\underbrace{\left(\begin{array}{ccc}\frac{1}{3}&0&0\\0&\frac{1}{3}&0\\0&0&-\frac{2}{3}\end{array}\right)}_{SU(3)_R}.
\end{equation}
\vspace{,25in}Tabulating the coefficients we have Table III.

\begin{center}
\bigskip
\begin{tabular}{|l|l|l|l|l|}
\hline
${\hat a}$ & ${\hat b}$ & ${\hat c}$ & ${\hat d}$ & type \\ \hline\hline

-1/6 & 1/6 & 1/2 & -1/6 & TR \\ \hline
-1/6 & $7/6$ & 0 & 1/3 & V \\ \hline
\end{tabular}

\bigskip

Table III: Class (iii) models.
\end{center}
\bigskip
\bigskip

As the first case is the TR model again we need only consider the second. Here the fermions are

 \begin{equation}
(\bar{4},1,3) \rightarrow  \underline{(3,{2})_{1/6}}+ (3,1)_{-4/3}+ (3,1)_{1}.
\end{equation}
  \begin{equation}
(1,3,\bar{3}) \rightarrow  (\bar{3},1)_{1/3}+\underline{(\bar{3},1)_{1/3}}+ \underline{(\bar{3},1)_{-2/3}}.
\end{equation}
 \begin{eqnarray}
(4,\bar{3},1) \rightarrow    \underline{(1,2)_{-1/2}}+(1,2)_{-1/2}+(1,2)_{1/2}\nonumber\\
~~~~~~+(1,1)_{-2/3}+(1,1)_{-2/3}+\underline{(1,1)_{1}}\nonumber\\
~~~~~~+(1,1)_{-1/3}+(1,1)_{-1/3}+(1,1)_{2/3}.
\end{eqnarray}
Multiple choices for selecting family components exist here of which we have underlined one possibility.


\section{The Higgs sector}
In this section, we will present a detailed analysis of the Higgs sector for the cases 
where the intermediate gauge group is the 422 (PS) group, the flipped 422 group and the 
trinification group. 
\section {Pati-Salam Models}


Consider the symmetry breaking chain
\[SU(4)\times SU(3)_A \times SU(3)_B \rightarrow SU(4) \times SU(2)_L \times SU(2)_R \rightarrow \frac{SU(3)_C \times U(1)_Y}{\mathbb{Z}_3}\times SU(2)_W,\]
which takes us from the 334 model to the standard model via the PS model, where we take
\[SU(2)_W=SU(2)_L \subset SU(3)_A,\]
\[SU(2)_R \subset SU(3)_B,\]
and
\[SU(3)_C \subset SU(4).\]
The weak hypercharge in this model is
\[Y_{PS}=\underbrace{\begin{pmatrix}\frac{1}{6}&0&0&0\\0&\frac{1}{6}&0&0\\0&0&\frac{1}{6}&0\\0&0&0&-\frac{1}{2}\end{pmatrix}}_{SU(4)}+\underbrace{\begin{pmatrix}\frac{1}{2}&0\\0&-\frac{1}{2}\end{pmatrix}}_{SU(2)_R}=\underbrace{\begin{pmatrix}\frac{1}{6}&0&0&0\\0&\frac{1}{6}&0&0\\0&0&\frac{1}{6}&0\\0&0&0&-\frac{1}{2}\end{pmatrix}}_{SU(4)}+\underbrace{\begin{pmatrix}\frac{1}{2}&0&0\\0&0&0\\0&0&-\frac{1}{2}\end{pmatrix}}_{SU(3)_B},\]
and is of PS type.

The fact that the SM gauge group is modded out by $\mathbb{Z}_3$ instead of the 
usual $\mathbb{Z}_6$ \cite{f4}
means that we can have electric charges which are multiples of $\frac{1}{2}$. At distances smaller than the QCD confinement scale, we can also have electric charges which are multiples of $\frac{1}{6}$, but at larger distances, we only find electric charges coming in multiples of $\frac{1}{2}$. However, these fractional charges are attached to particles with GUT scale\footnote{The GUT scale is defined to be the energy scale at which the GUT gauge symmetry is spontaneously broken.} masses.
Monopoles form at the PS breaking scale.

\subsection{Model PS$\alpha$}
\subsubsection{Fermion content}
The fermion content at the 334 level is
$3[(4,\bar{3},1)_{433}\oplus (\bar{4},1,3)_{433}\oplus (1,3,\bar{3})_{433}\oplus (1,3,1)_{433}\oplus (1,1,\bar{3})_{433}]$ \cite{f5}
with decomposition of the representations under the PS group given by
\[(4,\bar{3},1)_{433}\rightarrow (4,2,1)_{PS} \oplus (4,1,1)_{PS}.\]
\[(\bar{4},1,3)_{433}\rightarrow (\bar{4},1,2)_{PS}\oplus (\bar{4},1,1)_{PS}.\]
\[(1,3,\bar{3})_{433}\rightarrow (1,2,2)_{PS} \oplus (1,2,1)_{PS} \oplus (1,1,2)_{PS} \oplus (1,1,1)_{PS}.\]
\[(1,3,1)_{433}\rightarrow (1,2,1)_{PS} \oplus (1,1,1)_{PS}.\]
\[(1,1,\bar{3})_{433}\rightarrow (1,1,2)_{PS} \oplus (1,1,1)_{PS}.\]
We find the SM fermions (plus a left handed antineutrino) are contained within $(4,2,1)_{PS}$ and $(\bar{4},1,2)_{PS}$. All the other fermions are ``exotic'' and must be heavy enough to have escaped detection.




\subsubsection{Yukawa couplings}

Our goal is to make the exotic fermions vectorlike. In general, this can be done either at the 433 breaking scale or at the PS breaking scale. Since all the PS representations of the fermion fields other than the Standard Model + left handed antineutrino either come in conjugate pairs or are real representations, it is much more economical to arrange for all the pairings to occur at the 433 breaking scale.
Here, we assume that the pairings result from Yukawa terms after the Higgs field(s) acquire VEVs. Pairings are also possible using nonrenormalizable couplings \cite{f6}
or other means like dynamical symmetry breaking.

Pairing  $(4,1,1)_{PS}$ with $(\bar{4},1,1)_{PS}$ requires either \nolinebreak[3]{$\langle(15,3,\bar{3})_{H 433}\rangle$} \cite{f7}
or {\nolinebreak$\langle(1,3,\bar{3})\rangle_{H 433}$}. The choice of $15_{SU(4)}$ and $1_{SU(4)}$ comes about because they are the only two SU(4)-representations $\rho$ for which there exists a nonzero intertwiner\footnote{An intertwiner is a 433-invariant linear map from a representation into another.} from $\rho \otimes 4_{SU(4)} \otimes \bar{4}_{SU(4)}$ to the singlet representation where the interaction terms are gauge invariant. The VEV lies along the $(1,1,1)_{H PS}$ component of the Higgs field. The second choice is simpler, but the first choice works just as well, at least when it comes to Yukawa couplings. However, as we will see later, a $(1,3,\bar{3})_{H 433}$ Higgs is necessary for other reasons. So, the first choice may be an unnecessary complication. In summary, we choose either
\[\langle1,3,\bar{3}\rangle_{H 433}(4,\bar{3},1)_{433}(\bar{4},1,3)_{433},\]
or 
\[\langle15,3,\bar{3}\rangle_{H 433}(4,\bar{3},1)_{433}(\bar{4},1,3)_{433}.\]
$(1,2,2)_{PS}$ is real and the only $(1,2,2)_{PS}$'s come from $(1,3,\bar{3})_{433}$. So, the pairing has to be Majorana. Hence,
\[\langle(1,3,\bar{3})_{H 433}\rangle(1,3,\bar{3})_{433}(1,3,\bar{3})_{433}\]
is the only choice \cite{f8}.

Now let's move on to the $(1,2,1)_{PS}$ fermions. Neither
\[\langle(1,3,6)_{H 433}\rangle(1,3,\bar{3})_{433}(1,3,\bar{3})_{433}\]
nor
\[\langle(1,3,1)_{H 433}\rangle(1,3,1)_{433}(1,3,1)_{433}\]
are satisfactory because they lead to antisymmetric mass self-couplings of the fermion 
fields and with three generations, this still leaves us with some leftover massless 
fermions. Even if we have more than one such Higgs field with different Yukawa coupling 
constants between the generations, we are still left with massless fermions since the sum 
of two antisymmetric matrices is still an antisymmetric matrix which has a zero eigenvalue 
if its dimension is odd. It is not that we cannot add such couplings but that with such 
couplings, we still need to add the other Yukawa couplings to pair up all the exotics 
anyway and so, the addition of such couplings is an unnecessary complication, unless they 
are required to provide useful mass relations.
This leaves us with the choices that avoid asymmetric mass matrices, i.e., we require
\[\langle(1,3,3)\rangle_{H 433}(1,3,\bar{3})_{433}(1,3,1)_{433}\]
to give masses to the $(1,2,1)_{PS}$ fermions. Similarly, for the $(1,1,2)_{PS}$ fermions,
 we need the coupling
\[\langle(1,\bar{3},\bar{3})\rangle_{H 433}(1,3,\bar{3})_{433}(1,1,\bar{3})_{433}.\]

Finally, we are left to deal with the $(1,1,1)_{PS}$ fermions. They leave us with the 
greatest degree of freedom. However, since they do not couple to the PS gauge fields, 
(and as a consequence the SM gauge fields), and as all the SM fermions are located within $(4,\bar{3},1)_{433}$ and $(\bar{4},1,3)_{433}$ and as these singlets are located within the $SU(4)$-neutral representations, we have not added a single Yukawa coupling which mixes the $SU(4)$-charged fermions with the $SU(4)$-neutral fermions, it is unnecessary to pair them up.



Note that all of the Higgs fields introduced previously with the possible exception of the 
optional $(15,3,\bar{3})_{H 433}$ are $SU(4)$-singlets. This means that we need additional 
Higgs fields which are not $SU(4)$-singlets to break PS down to SM. These Higgs fields will acquire VEVs at a lower energy scale compared to the $SU(4)$-neutral Higgs fields.

Since the electroweak doublets couple some SM components of $(4,\bar{3},1)_{433}$ to some 
SM components of $(\bar{4},1,3)_{433}$, they have to lie in either $(15,3,\bar{3})_{H 433}$ 
or $(1,3,\bar{3})_{H 433}$ or some linear combination. As $(15,3,\bar{3})_{H 433}$ is not 
really needed, the second choice is the simplest one. (We still need to break $SU(4)$ to 
$SU(3)\times U(1)$, and the most efficient choice for this is $(15,1,1)_{H 433}$.) If  
only one $(1,3,\bar{3})_{H 433}$ field exists, this will lead to mass relations between 
the up-type and down-type quarks which are not observed. So, there has to be at least two such fields.

In a full phenomenological model, we will have to make sure that all the Higgs fields pair up except for the electroweak doublets (unless there is some mechanism making the additional Higgs fields phenomenologically harmless) with the possible exception of some SM-singlet Higgs fields (or even electroweak triplets, but then we might have to worry about protecting the tiny mass of the left-handed neutrino), but we will not work this out here. Higgs fields with $SU(3)_C$-color are likely to cause rapid proton decay.

Lastly, the left-handed antineutrino has to acquire a Majorana mass. There are number of ways to do this. For instance,
\[\langle(10,1,\bar{6})_{H 433}\rangle(\bar{4},1,3)_{433}(\bar{4},1,3)_{433}\]
is one of them. A Yukawa coupling with a Higgs VEV associated with the PS breaking scale of the left-handed antineutrino to one of the vector exotics which already has a mass associated with the 433 breaking scale is another. Or we can even add singlet fermion fields $(1,1,1)_{433}$ and pair up the left-handed antineutrino with it.
Any of these methods will lead to a small seesaw mass for the left-handed neutrino.

\subsection{Model PS$\beta$}

\subsubsection{Fermion content}
Now consider the 334 model with pure bifundamental fermions, 
$3(4,\bar{3},1)_{433}\oplus 3(\bar{4},1,3)_{433}\oplus 4(1,3,\bar{3})_{433}$
and therefore family unification. These fermions decompose via
\[(4,\bar{3},1)_{433}\rightarrow (4,2,1)_{PS} \oplus (4,1,1)_{PS},\]
\[(\bar{4},1,3)_{433}\rightarrow (\bar{4},1,2)_{PS}\oplus (\bar{4},1,1)_{PS},\]
and
\[(1,3,\bar{3})_{433}\rightarrow (1,2,2)_{PS} \oplus (1,2,1)_{PS} \oplus (1,1,2)_{PS} \oplus (1,1,1)_{PS}.\]


\subsubsection{Yukawa couplings}
The symmetry breaking can be analysed by using the same reasoning as in the previous model,
where we choose either
\[\langle(1,3,\bar{3})_{H 433}\rangle(4,\bar{3},1)_{433}(\bar{4},1,3)_{433},\]
or
\[\langle(15,3,\bar{3})_{H 433}\rangle(4,\bar{3},1)_{433}(\bar{4},1,3)_{433},\]
to give masses to the $(4,1,1)_{PS}$ and $(\bar{4},1,1)_{PS}$.
The rest of the Yukawa terms are determined (after making the same assumptions as 
previously discussed) to be

\[
\begin{matrix}
\langle(1,3,\bar{3})_{H 433}\rangle(1,3,\bar{3})_{433}(1,3,\bar{3})_{433},\\
\langle(1,3,6)_{H 433}\rangle(1,3,\bar{3})_{433}(1,3,\bar{3})_{433},\\
\langle(1,\bar{6},\bar{3})_{H 433}\rangle(1,3,\bar{3})_{433}(1,3,\bar{3})_{433}.
\\
\end{matrix}
\] The fact that two of the couplings lead to antisymmetric mass matrices is no problem 
here since there are {\it four}  $(1,3,\bar{3})$s  and as before, we do not need to pair up $(1,1,1)_{PS}$.

The comment that additional Higgs fields are needed to break the PS gauge group also applies here. So do the comments 
about the electroweak Higgs doublets and giving the left-handed antineutrino a large Majorana mass. For instance, if 
we avoid the $\langle(15,3,\bar{3})_{H 433}\rangle$ above, then we need to include a $\langle(15,1,1)_{H 433}\rangle $ to break $SU(4)$.

\section{Model FPSa}

The symmetry breaking chain is \cite{f9}
\[SU(4)\times SU(3)_A \times SU(3)_B \rightarrow SU(4) \times SU(2)_L \times \frac{SU(2)_R \times U(1)_X}{\mathbb{Z}_2} \rightarrow \frac{SU(3)_C \times U(1)_Y}{\mathbb{Z}_3} \times SU(2)_W,\]
where
\[SU(2)_W =SU(2)_L \subset SU(3)_A,\]
\[SU(2)_R \subset SU(3)_B,\]
and
\[SU(3)_C \subset SU(4).\] Defining the generator
\[Z=\underbrace{\begin{pmatrix}1&0&0\\0&1&0\\0&0&-2\end{pmatrix}}_{SU(3)_B}\]
we can write the hypercharge in the form
\[Y=\underbrace{\begin{pmatrix}\frac{1}{6}&0&0&0\\0&\frac{1}{6}&0&0\\0&0&\frac{1}{6}&0\\0&0&0&-\frac{1}{2}\end{pmatrix}}_{SU(4)}+\underbrace{\begin{pmatrix}\frac{1}{4}&0\\0&-\frac{1}{4}\end{pmatrix}}_{SU(2)_R}+\frac{Z}{4}=\underbrace{\begin{pmatrix}\frac{1}{6}&0&0&0\\0&\frac{1}{6}&0&0\\0&0&\frac{1}{6}&0\\0&0&0&-\frac{1}{2}\end{pmatrix}}_{SU(4)}+\underbrace{\begin{pmatrix}\frac{1}{2}&0&0\\0&0&0\\0&0&-\frac{1}{2}\end{pmatrix}}_{SU(3)_B}\]

The embedding of the SM gauge group within 433 is the same as in the unflipped PS model (which means that they are 
the same type of model in our classification). The matter sector, the Higgs sector, and the Yukawa couplings are also 
the same. The difference lies in the relative values of the different Higgs VEVs, which determines the intermediate 
gauge group. This is not unlike the case of some $SO(10)$ models, for example, where despite having the same matter 
and Higgs sectors and Yukawa couplings, the intermediate gauge group may be 622 or $SU(5)$, depending upon the 
relative scale of $\langle 54_{H SO(10)} \rangle$ and
$\langle 16_{H SO(10)}\rangle/\langle\overline{16}_{H SO(10)}\rangle$.

At distances smaller than the QCD confinement scale, we can again have electric charges which are multiples of 
$\frac{1}{6}$, but at larger distances, we only find electric charges in multiples of $\frac{1}{2}$. However, these fractional charges are for particles with GUT scale masses.
Again, monopoles form at the 433 breaking scale.

\subsection{Model FPSa$\alpha$}

Again we must consider fermions with family replication
\[3[(4,\bar{3},1)_{433}\oplus (\bar{4},1,3)_{433}\oplus (1,3,\bar{3})_{433}\oplus (1,3,1)_{433} \oplus (1,1,\bar{3})_{433}],\]
with decompositions
\[(4,\bar{3},1)_{433} \rightarrow (4,2,1)_{0 FPSA} \oplus (4,1,1)_{0 FPSA},\]
\[(\bar{4},1,3)_{433} \rightarrow (\bar{4},1,2)_{1 FPSA} \oplus (\bar{4},1,1)_{-2 FPSA},\]
\[(1,3,\bar{3})_{433} \rightarrow (1,2,2)_{-1 FPSA} \oplus (1,2,1)_{2 FPSA} \oplus (1,1,2)_{-1 FPSA} \oplus (1,1,1)_{2 FPSA},\]
\[(1,3,1)_{433} \rightarrow (1,2,1)_{0 FPSA} \oplus (1,1,1)_{0 FPSA},\]
and
\[(1,1,\bar{3})_{433} \rightarrow (1,1,2)_{-1 FPSA} \oplus (1,1,1)_{2 FPSA},\]
where we use the following array for definitions 
\[
\begin{array}{|cccc|}
(4,2,1)_{0 FPSA}&(\bar{4},1,2)_{1 FPSA}&(4,1,1)_{0 FPSA}&(\bar{4},1,1)_{-2 FPSA}\\
\hline\hline
  \begin{pmatrix}
    (3,2)_{\frac{1}{6}}&q\\
    (1,2)_{-\frac{1}{2}}&l
  \end{pmatrix}
  &
  \begin{pmatrix}
    (\bar{3},1)_{\frac{1}{3}}&d^c\\
    (\bar{3},1)_{-\frac{1}{6}}\\
    (1,1)_{1}&e^c\\
    (1,1)_{\frac{1}{2}}\\
  \end{pmatrix}
  &
  \begin{pmatrix}
    (3,1)_{\frac{1}{6}}\\
    (1,1)_{-\frac{1}{2}}
  \end{pmatrix}
  &
  \begin{pmatrix}
    (\bar{3},1)_{-\frac{2}{3}}&u^c\\
    (1,1)_{0}&\nu^c
  \end{pmatrix}
\end{array}
\]
The SM fermions are flipped, which is why this model is called a flipped PS model. Both $(4,1,2)_{1 FPSA}$ and $(\bar{4},1,1)_{0 FPSA}$ contain additional exotics. 


\subsubsection{Yukawa couplings}
Now either
\[\langle(1,3,\bar{3})_{H 433}\rangle(4,\bar{3},1)_{433}(\bar{4},1,3)_{433}\]
or
\[\langle(15,3,\bar{3})_{H 433}\rangle(4,\bar{3},1)_{433}(\bar{4},1,3)_{433}\]
is necessary to begin to carry out the SSB in this model.
The former Higgs fields decompose as 
\[(1,3,\bar{3})_{H 433}\rightarrow (1,2,2)_{-1 H FPSA} \oplus (1,2,1)_{2 H FPSA} \oplus (1,1,2)_{-1 H FPSA} \oplus (1,1,1)_{2 H FPSA},\]
which makes it clear that its VEVs are at the FPSA breaking scale, since none of its components are FPSA-singlets.

The Yukawa coupling responsible for pairing of exotics is

\[\langle(1,1,2)_{-1 H FPSA}\rangle(4,1,1)_{0 FPSA}(\bar{4},1,2)_{1 FPSA}\]
None of the  other Yukawa terms lead to further pairings until the electroweak breaking scale. The down-type and 
up-type electroweak Higgs doublets are contained within $(1,2,2)_{-1 H FPSA}$ and $(1,2,1)_{2 H FPSA}$ respectively.
The term $(1,2,2)_{-1 H FPSA}(4,2,1)_{0 FPSA}(\bar{4},1,2)_{1 FPSA}$ contains $(1,2)_{-\frac{1}{2} H}q d^c$ and $(1,2)_{-\frac{1}{2} H} l e^c,$ while $(1,2,1)_{2 H FPSA}(4,2,1)_{0 FPSA}(\bar{4},1,1)_{-2 FPSA}$ contains $(1,2)_{\frac{1}{2} H}q u^c$ and $(1,2)_{\frac{1}{2} H}l \nu^c.$

The decompositions of the other fermions are displayed below:
\[
\begin{array}{|ccc|}
(1,3,\bar{3})_{433}&(1,3,1)_{433}&(1,1,\bar{3})_{433}\\
\hline\hline
  \begin{pmatrix}
    (1,2)_{-\frac{1}{2}}&(1,1)_{-\frac{1}{2}}\\
    (1,2)_{0}&(1,1)_{0}\\
    (1,2)_{\frac{1}{2}}&(1,1)_{\frac{1}{2}}
  \end{pmatrix}
  &
  \begin{pmatrix}
    (1,2)_{0}&(1,1)_{0}
  \end{pmatrix}
  &
  \begin{pmatrix}
    (1,1)_{-\frac{1}{2}}\\
    (1,1)_{0}\\
    (1,1)_{\frac{1}{2}}
  \end{pmatrix}\\
\end{array} ~.
\] The $(1,2)_{\frac{1}{2}}$ of $(1,3,\bar{3})_{433}$ has to pair up with the $(1,2)_{-\frac{1}{2}}$ of $(1,3,\bar{3})_{433}$. This is done using the coupling
\[\langle(1,3,\bar{3})_{H 433}\rangle(1,3,\bar{3})_{433}(1,3,\bar{3})_{433}.\]

Since there are three generations of $(1,3,\bar{3})_{433}$ and the self-pairing of $(1,2)_{0}$ is antisymmetric since 
it is a pseudoreal representation, we have to pair the $(1,2)_{0}$ of $(1,3,\bar{3})_{433}$ with the 
$(1,2)_{0}$ of $(1,3,1)_{433}$. So, the coupling
\[\langle(1,3,3)_{H 433}\rangle(1,3,\bar{3})_{433}(1,3,1)_{433}\]
is required.

To pair up $(1,1)_{\frac{1}{2}}$ and $(1,1)_{-\frac{1}{2}}$, 
the following two couplings are unnecessary
\[\langle(1,\bar{6},\bar{3})_{H 433}\rangle(1,3,\bar{3})_{433}(1,3,\bar{3})_{433}\]
\[\langle(1,1,\bar{3})_{H 433}\rangle(1,1,\bar{3})_{433}(1,1,\bar{3})_{433}\]
for the same reason of antisymmetry and an odd number of generations.

Finally, we are left with the choice
\[\langle(1,\bar{3},\bar{3})_{H 433}\rangle(1,3,\bar{3})_{433}(1,1,\bar{3})_{433}.\]
There is no reason to pair up the SM-singlet fermions here either,
so we are left with the same four Yukawa coupling terms as in the previous model.

Since $SU(4)$ must be broken, there have to be Higgs fields which transform under  $SU(4)$ nontrivially and acquire VEVs at the FPSA breaking scale. Unless there is some principle (symmetry or otherwise) \cite{f10}
or some dynamical mechanism preventing us, we might expect to have Yukawa (or nonrenormalizable) couplings involving this Higgs field, a $SU(4)$-charged fermion field and an $SU(4)$-neutral fermion field. An example is
\[\langle(\bar{4},1,3)_{H 433}\rangle(4,\bar{3},1)_{433}(1,3,\bar{3})_{433}.\]
This can cause some further flipping so that the standard model fermions are really linear combinations of some 
components of $(4,\bar{3},1)_{433}$ and some components of $(1,3,\bar{3})_{433}$. This modifies the electroweak 
Yukawa couplings so that we might get away with only one $(1,3,\bar{3})_{H 433}$ Higgs field, instead of two, and 
still get the right mass relations among the fermions. But in that case, we definitely need to pair up all the 
SM-singlet fermions or otherwise, there will be observable mixings. This is unlike the case of the unflipped PS 
model, because the leptons in the $(1,3,\bar{3})_{433}$ (and also $(1,3,1)_{433}$ and $(1,1,\bar{3})_{433}$ in some 
models) acquire 433 breaking scale masses and a PS scale cross-coupling between $(4,\bar{3},1)_{433}$ or 
$(\bar{4},1,3)_{433}$ with $(1,3,\bar{3})_{433}$ will only change the SM lepton eigenstates by a tiny angle.

\subsection{Model FPSa$\beta$}
This construction is similar to the previous model, so we shall not go through the details here.

\section {Model FPSb}
Next consider the symmetry breaking chain
\[SU(4)\times SU(3)_A \times SU(3)_B \rightarrow SU(4) \times SU(2)_L \times \frac{SU(2)_R \times U(1)_X}{\mathbb{Z}_2} \rightarrow \frac{SU(3)_C \times U(1)_Y}{\mathbb{Z}_3} \times SU(2)_W,\]
where
\[SU(2)_L \subset SU(3)_A, \]
and
\[SU(2)_R \subset SU(3)_B. \]
If we let
\[U=\underbrace{\begin{pmatrix}-2&0&0\\0&1&0\\0&0&1\end{pmatrix}}_{SU(3)_B},\]
be the generator of $U(1)_X$, then we can write the hypercharge as
\[Y=\underbrace{\begin{pmatrix}\frac{1}{6}&0&0&0\\0&\frac{1}{6}&0&0\\0&0&\frac{1}{6}&0\\0&0&0&-\frac{1}{2}\end{pmatrix}}_{SU(4)}+\underbrace{\begin{pmatrix}\frac{1}{4}&0\\0&-\frac{1}{4}\end{pmatrix}}_{SU(2)_R}-\frac{U}{4}=\underbrace{\begin{pmatrix}\frac{1}{6}&0&0&0\\0&\frac{1}{6}&0&0\\0&0&\frac{1}{6}&0\\0&0&0&-\frac{1}{2}\end{pmatrix}}_{SU(4)}+\underbrace{\begin{pmatrix}\frac{1}{2}&0&0\\0&0&0\\0&0&-\frac{1}{2}\end{pmatrix}}_{SU(3)_B}.\]
This model has similarities with the previous models (both flipped I and flipped II),  but the embedding of the intermediate group is different, so the difference between this model and the previous two models lies in the relative scale of the Higgs VEVs.

The relevant differences lie in the nontrivial $SU(4)$ irreducible representations displayed below:

\[
\begin{array}{|cccc|}
(4,2,1)_{0 FPSB}&(\bar{4},1,2)_{1 FPSB}&(4,1,1)_{0 FPSB}&(\bar{4},1,1)_{-2 FPSB}\\
\hline\hline
  \begin{pmatrix}
    (3,2)_{\frac{1}{6}}&q\\
    (1,2)_{-\frac{1}{2}}&l
  \end{pmatrix}
  &
  \begin{pmatrix}
    (\bar{3},1)_{-\frac{1}{6}}\\
    (\bar{3},1)_{-\frac{2}{3}}&u^c\\
    (1,1)_{\frac{1}{2}}\\
    (1,1)_{0}&\nu^c
  \end{pmatrix}
  &
  \begin{pmatrix}
    (3,1)_{\frac{1}{6}}\\
    (1,1)_{-\frac{1}{2}}
  \end{pmatrix}
  &
  \begin{pmatrix}
    (\bar{3},1)_{\frac{1}{3}}&d^c\\
    (1,1)_{1}&e^c
  \end{pmatrix}
  \\
\end{array}
\]
The up-type electroweak Higgs doublet is contained within $(1,2,2)_{-1 H FPSB}$ and the down-type Higgs within $(1,2,1)_{2 FPSB}$, which is the reverse of the case with FPSA.

\subsection{Direct breaking}

It is also certainly possible that all the GUT Higgs VEVs have the same order of magnitude. In that case, it is more 
appropriate to say that 433 breaks down directly to the SM gauge group.

\section{Trinification}
Finally, we return to the trinification case where we will explore the variant of the model with SSB
\[SU(4)\times SU(3)_A \times SU(3)_B \rightarrow SU(3)_C \times SU(3)_L \times SU(3)_R \rightarrow SU(3)_C \times \frac{SU(2)_W \times U(1)_Y}{\mathbb{Z}_2}.\]
The $\mathbb{Z}_2$ modding means that the electric charge is quantized in multiples of 1/3 \cite{f11}
. Monopoles form at the TR breaking scale. We choose
the fermion content \\
$3[(4,\bar{3},1)\oplus (\bar{4},1,3)\oplus (1,3,\bar{3})\oplus (1,3,1)\oplus (1,1,\bar{3})].$

\subsection{Model TRa}
The first model has
\[SU(3)_C \subset SU(4),\]
\[SU(2)_W \subset SU(3)_L = SU(3)_A,\]
and
\[SU(3)_R = SU(3)_B,\]
with hypercharge
\[Y_{TRa}=\underbrace{\begin{pmatrix}-\frac{1}{6}&0&0\\0&-\frac{1}{6}&0\\0&0&\frac{1}{3}\end{pmatrix}}_{SU(3)_A}+\underbrace{\begin{pmatrix}\frac{1}{3}&0&0\\0&\frac{1}{3}&0\\0&0&-\frac{2}{3}\end{pmatrix}}_{SU(3)_B}.\]

The decompositions of the representations under TR are

\[(4,\bar{3},1)_{433}\rightarrow (3,\bar{3},1)_{TR} \oplus (1,\bar{3},1)_{TR},\]
\[(\bar{4},1,3)_{433}\rightarrow (\bar{3},1,3)_{TR} \oplus (1,1,3)_{TR},\]
\[(1,3,\bar{3})_{433}\rightarrow (1,3,\bar{3})_{TR},\]
\[(1,3,1)_{433}\rightarrow (1,3,1)_{TR},\]
and
\[(1,1,\bar{3})_{433}\rightarrow (1,1,\bar{3})_{TR}.\]



\subsubsection{Yukawa couplings}

The fractionally charged fermions are located within the TR-fundamental fermions fields but not the TR-bifundamental fermions fields. So, we have to pair up the TR-fundamentals with TR-fundamentals. This can be done with a 433 breaking term
\[\langle(\bar{4},1,1)_{H 433}\rangle(4,\bar{3},1)_{433}(1,3,1)_{433}\]
which takes care of $(1,\bar{3},1)_{TR}$ and $(1,3,1)_{TR}$. If we wish to be exhaustive, we should not overlook the possibility of using a VEV for an $SU(3)_A$-octet 
\[\langle(\bar{4},8,1)_{H 433}\rangle(4,\bar{3},1)_{433}(1,3,1)_{433}.\]
This Higgs field can only acquire a VEV at the TR breaking scale, with the corresponding pairing
\[\langle(1,8,1)_{H TR}\rangle(1,\bar{3},1)_{TR}(1,3,1)_{TR},\]
which gives a different mass relation between the exotic fermions from the previous choice.

Similarly, we can  have either
\[\langle(4,1,1)_{H 433}\rangle(\bar{4},1,3)_{433}(1,1,\bar{3})_{433},\]
or
\[\langle(4,1,8)_{H 433}\rangle(\bar{4},1,3)_{433}(1,1,\bar{3})_{433}\]
with a different mass relation in each case.

In trinification, we need more than one $(1,3,\bar{3})_{H TR}$ Higgs field. The other Yukawa couplings terms are
\[\langle(1,3,\bar{3})_{H 433}\rangle(4,\bar{3},1)_{433}(\bar{4},1,3)_{433},\]
( $(15,3,\bar{3})_{H 433}$ is also possible, but unnecessary since we already have $(1,3,\bar{3})_{H 433}$'s),
and
\[\langle(1,3,\bar{3})_{H 433}\rangle(1,3,\bar{3})_{433}(1,3,\bar{3})_{433}.\]
This is the only choice which pairs up the exotic leptons. However, it does not cause the SM-neutral fermions to pair up \cite{f12}.
A $(1,\bar{6},\bar{3})_{H 433}$ leads to antisymmetric mass matrix, but a $(1,\bar{6},6)_{H 433}$   can be responsible for giving large symmetric masses to the remaining exotics. 

\subsection{Model TRb}
Our second trinification model has 
\[SU(3)_C = SU(3)_A,\]
\[SU(2)_W \subset SU(3)_L = SU(3)_B,\]
and
\[SU(3)_R \subset SU(4),\]
with hypercharge
\[Y_{TRb}=\underbrace{\begin{pmatrix}\frac{1}{3}&0&0&0\\0&\frac{1}{3}&0&0\\0&0&-\frac{2}{3}&0\\0&0&0&0\end{pmatrix}}_{SU(4)}+\underbrace{\begin{pmatrix}-\frac{1}{6}&0&0\\0&-\frac{1}{6}&0\\0&0&\frac{1}{3}\end{pmatrix}}_{SU(3)_B}.\]


\subsubsection{Yukawa couplings}
The appropriate Yukawa couplings terms are
\[
\begin{matrix}
\langle(\bar{4},1,1)_{H 433}\rangle(4,\bar{3},1)_{433}(1,3,1)_{433},\\
\langle(4,1,1)_{H 433}\rangle(\bar{4},1,3)_{433}(1,1,\bar{3})_{433},\\
\langle(\bar{4},1,3)_{H 433}\rangle(4,\bar{3},1)_{433}(1,3,\bar{3})_{433},\\
\langle(6,1,3)_{H 433}\rangle(\bar{4},1,3)_{433}(\bar{4},1,3)_{433}.\\
\end{matrix}
\]
As before, the SM-singlet fermions may be paired using a term of the form
\[\langle(10,1,\bar{6})_{H 433}\rangle(\bar{4},1,3)_{433}(\bar{4},1,3)_{433}.\]

\subsection{Model TRb'}
The third model has some similarities with Model TRa. Here we have
\[SU(3)_C=SU(3)_B,\]
\[SU(2)_W \subset SU(3)_L \subset SU(4),\]
and
\[SU(3)_R =SU(3)_A,\]
with hypercharge
\[Y_{TRb'}=\underbrace{\begin{pmatrix}-\frac{1}{6}&0&0&0\\0&-\frac{1}{6}&0&0\\0&0&\frac{1}{3}&0\\0&0&0&0\end{pmatrix}}_{SU(4)}+\underbrace{\begin{pmatrix}\frac{1}{3}&0&0\\0&\frac{1}{3}&0\\0&0&-\frac{2}{3}\end{pmatrix}}_{SU(3)_A}.\]


\subsubsection{Yukawa couplings}
The necessary Yukawa couplings terms are
\[
\begin{matrix}
\langle(\bar{4},1,1)_{H 433}\rangle(4,\bar{3},1)_{433}(1,3,1)_{433},\\
\langle(4,1,1)_{H 433}\rangle(\bar{4},1,3)_{433}(1,1,\bar{3})_{433},\\
\langle(4,\bar{3},1)_{H 433}\rangle(\bar{4},1,3)_{433}(1,3,\bar{3})_{433},\\
\langle(6,\bar{3},1)_{H 433}\rangle(4,\bar{3},1)_{433}(4,\bar{3},1)_{433},\\
\end{matrix}
\]
and the SM-singlet fermions may be paired up via
\[\langle(\overline{10},6,1)_{H 433}\rangle(4,\bar{3},1)_{433}(4,\bar{3},1)_{433}.\]

For the remaining models in our classication, we have not been able to find a suitable choice of additional 
``exotic'' fermion fields to add to make {\bf all} the exotic fermions vectorlike at the GUT scale. While this 
certainly does not mean that such a combination is not possible, it probably means that any such combination would 
have to be fairly complicated.

\section{Discussion}
Perhaps the most unusual and interesting new models found here are those where some of the extra fermions have  $X$ 
charges. We can generate other models similar to the X model by extending the gauge group of the PS or TR model. 
To see this let one of the $SU(N)$s be extended to $SU(N+2)$ in  a way that the families have no charge under the 
new diagonal generator $X'=diag(0,0,...0,1,-1)$
of $SU(N+2)$. For example, let $G_{TR} \mbox{ be replaced by } SU(3)\times SU(3)\times SU(5)$, where we then break this symmetry to $G_{TR}\times U(1)_{X'}$. Then the TR fermions which must be extended to
 \begin{equation}
5(3,\bar{3},1) +3(1,3,\bar{5})+3(\bar{5},1,3)  
\end{equation}
under $G_{TR}\rightarrow SU(3)\times SU(3)\times SU(5)$,  reduce to 
 \begin{eqnarray}
5(3,\bar{3},1)_0 & + & 3(1,3,\bar{3})_0 + 3(\bar{3},1,3)_0\nonumber\\
& + & (1,3,1)_{x'}+(\bar{3},1,1)_{x'}\nonumber\\
& + & (1,3,1)_{-x'}+(\bar{3},1,1)_{-x'}.
\end{eqnarray}
This model has some interest in its own right since family unification is required if we only allow bifundamentals 
(However, problems can arise in this model with respect to giving heavy masses to the exotics.). As planned, the three families have no $X'$ charge. Therefore, the breaking of $U(1)_{X'}$ is not tied to the charge operator for standard families, nor is the $X'$ charge {\it a priori} quantized for the extra fermions. The X model itself is somewhat more interesting since it is not one of these extensions.
$U(1)_X$ is an integral part of the 334 model and is interwoven into $SU(4)\times SU(3)\times SU(3)$ in a nonfactorizable way. As we have seen, there are a number of choices for $x$ that can be used to flip the $X$ model. Likewise, this can be done for $X'$ models.

The spontaneous symmetry breaking of $SU(4)\times SU(3)\times SU(3)$ to  models 
that contain the TR or PS model is straightforward. A Higgs multiplet $(4,1,1)_H+h.c.$ can be used to break directly 
to the $SU^3(3)$ of the TR model. Further breaking proceeds as in the TR model. Likewise a $(1,3,1)_H+h.c.$ and a 
$(1,1,3)_H+h.c.$ can be used to break $SU(4)\times SU(3)\times SU(3)$ to the PS group $SU(4)\times SU(2)\times SU(2)$. 
Further breaking proceeds as in the PS model. Hence, we only need to concern ourselves with the 334 that is not 
equivalent to the TR or PS models. The SSB to any of the models listed can always be achieved, given sufficient 
freedom in the Higgs sector. It would be interesting if we could derive any of the 334 type models from an 
orbifolded $AdS/CFT$ theory, as this would strongly restrict both the fermion and scalar content of the theory, 
i.e., adjoints and bifundamentals only.


In the PS and TR models, several symmetry breaking scales can be 
associated with magnetic monopoles.  In this context it is important that the
gauge boson mediated proton decay is absent in
these models, but proton decay can still proceed through Higgs (Higgsino if the model is supersymmetrized)
exchange, as well as via higher dimension operators.
It is relatively straightforward to construct 
models based on $G_{PS}$ and $G_{TR}$ where proton
decay is forbidden, as a 
consequence, say of  an `accidental' baryon number symmetry.
This allows the possibility that $G_{PS}$ and $G_{TR}$ 
could be broken at scales far below the conventional
grand unification scale $M_{GUT}\sim 10^{16} GeV$.
An example based on D-branes in Type I string theory was
provided for the $G_{PS}$ symmetry
\cite{Leontaris:2000hh}, where with the
standard embedding of $SU(3)_{C} \times SU(2) \times U(1)$,
the symmetry breaking scale of $G_{PS}$ becomes
$M_{PS}\sim 10^{12} - 10^{13}$ GeV, with the corresponding
string scale $\stackrel{_>}{_\sim} M_{PS}$.  
Thus, monopoles with mass $\sim 10^{13} - 10^{14}$ GeV
are expected in
this class of models.  
An even more suggestive result is given by the PS type model based on
$CFT$
obtained from orbifolded type $IIB$ strings 
\cite{Frampton:2000zy},
\cite{Frampton:2000mq}.
Here the
unification is in the $100$ $TeV$ range,  and other intriguing phenomenology possibilities appear, e.g., $sin^{2}\theta_W$ can be 
approximately $.23$, etc.
\cite{Frampton:2001xh}. 
Analogous considerations should apply to
the trinification scheme and, by extension, to the
gauge symmetry of special interest here $SU(4) \times SU(3) \times SU(3)$. Assuming for instance that a 334 model could be derived from an orbifolded $AdS \times S^5$, the multiply
charged monopoles of the theory will have mass as small as $M\sim10^{7}$ GeV, which is in the preferred 
range of interest if they are to be candidates for
high
energy cosmic ray primaries
\cite{Kephart:1996bi},
\cite{Wick:2000yc};
a more detailed study of mass scales would require an RG analysis for each model.
We expect
the
334-model to have a similar unification scale with resulting exotic
(fractionally charged) leptons and/or hadrons, and we expect their masses
to be
near this unification scale, so they are also of interest as dark matter
candidates
\cite{Albuquerque:2000rk},
\cite{Chung:2001cb}.  


Given that monopoles of mass of order $\sim 10^{13} - 10^{14}$ GeV
(or perhaps even much lighter) can arise in realistic
models, it is important to ask:  Can these primordial monopoles 
survive inflation?
A non-supersymmetric 
inflationary scenario which dilutes but does not completely
wash away intermediate mass monopoles was developed in ref
\cite{Lazarides}.
The D-brane scenario discussed above gives rise to 
non-supersymmetric $SU(4)_{C} \times SU(2) \times SU(2)$,
so the discussion in ref
\cite{Lazarides}  may be relevant.  The monopole
flux can be reduced to close to the Parker bound 
$10^{-16} cm^{-2} s^{-1} sr^{-1}$. In the orbifolded scheme, the SSB scale 
 where the monopoles get their masses can be below 
the inflation scale. Hence, the monopoles can exist in interesting densities (near the Parker bound)
depending on details of the SSB phase transitions.
For the supersymmetric case, dilution of monopoles can
be achieved by thermal inflation
\cite{Pana},
\cite{Lyth}
 followed by entropy production.
A scenario in which thermal inflation is associated with the
breaking of the $U(1)$ axion symmetry was recently developed in
ref \cite{Laz}.

In summary, the $SU(4) \times SU(3) \times SU(3)$ models we are
advocating provides a natural family unification while avoiding proton decay and
giving rise to both (exotic) fractionally charged
color singlets and corresponding multiply charged magnetic
monopoles
\cite{King:1998ia} with densities compatible with the Parker bound, and
with masses 
perhaps as light
as $\sim 10^7$ GeV,
(Note that in $SU(5)$ the lightest monopole has mass of
$\sim 10^{17}$ GeV, and carries one unit of magnetic charge
\cite{Daniel:1980yz}.)
The exotic states are heavy (greater than a few TeV), but may be in the range 
explored by accelerators in the next decade. 

\vspace{.5in}
\noindent {\it Acknowledgments:} TWK thanks the Bartol Research Institute at the University of Delaware and the 
Aspen Center for Physics for hospitality while this work was in progress. This work was supported by the US 
Department of Energy under Grants No. DE-FG05-85ER40226 (TWK) and by DE-FG02-91ER40626 (QS).

\section{}
\subsection{}
\subsubsection{}


\begin{thebibliography}{99}
\bibitem{f1}
We could easily extend this discussion to models with more $SU(N)$ factors.
\bibitem{f2}
Recall that three standard families of chiral fermions 
at the EW breaking scale are in the representations: $3[(3,2)_\frac{1}{6}\oplus
(\bar{3},1)_{-\frac{2}{3}}\oplus(\bar{3},1)_{\frac{1}{3}}
\oplus(1,2)_{-\frac{1}{2}}\oplus(1,1)_1]$.
If we include right handed neutrino, they are in neutral singlets $ (1,1)_0$.
\bibitem{footnote}
An example of such a model obtained from orbifolding 
might be $S^{5}/\Gamma $, 
where $\Gamma$ contains 3 and 4 dimensional irreducible representations and
we
choose $N=1$.
\bibitem{f3}
It is standard practice to normalize all $U(1)$ generators 
 $\Lambda$ such that $Tr[\Lambda^2]=2$ but we usually find it 
 more convenient to drop the overall normalization factors as will be 
 either stated or clear from context.
\bibitem{fx}
Technically speaking, we should not be using the vacuum since 
hypercharge is spontanously broken in the vacuum. 
We should be using some thermal state in equilibrium at a 
temperature above the electroweak phase transition.
\bibitem{f4}
This can be traced to the fact that we are breaking down 
from $SU(4)\times SU(2)_L \times SU(2)_R$ instead of 
$\frac{SU(4)\times SU(2)_L \times SU(2)_R}{\mathbb{Z}_2}$
\bibitem{f5}
In the rest of this draft, we will append a subscript to each representation 
indicating the Lie group it is a representation of to reduce any potential ambiguities 
that may arise.
\bibitem{f6}
This might actually allow us to reduce the number of Higgs fields.
\bibitem{f7}
A subscript of H indicates that it is a Higgs field. The absence 
of such a subscript indicates that it is a fermion field.
\bibitem{f8}
Let's say an SU(3) symmetry breaks down to SU(2) and any Higgs VEV has to be invariant under SU(2). Let's also say that we have two SU(3)-triplet fermions. The SU(3)-triplets decompose into an SU(2)-doublet and an SU(2)-singlet. Suppose we wish to pair up the SU(2)-doublets using a Yukawa coupling to a Higgs field. Then a $\bar{6}_{H SU(3)}$ Higgs does not work because $\langle\bar{6}_{H SU(3)}\rangle3_{SU(3)}3_{SU(3)}$ only pairs up the SU(2)-singlets but not the SU(2)-doublets. So we have to go for the $\langle3_{H SU(3)}\rangle3_{SU(3)}3_{SU(3)}$ coupling instead, which only pairs up the SU(2)-doublets but not the SU(2)-singlets. Conversely, if we only wish to pair up the SU(2)-singlets but not the SU(2)-doublets, we have to choose $\bar{6}_{H SU(3)}$ and if we wish to pair up both the SU(2)-doublets and the SU(2)-singlets, we have to include both Higgs fields.
\bibitem{f9}
This will be abbreviated as FPSA.
\bibitem{f10}
There could be some additional symmetries preventing such couplings, or maybe it is the adjoint $15_{SU(4)}$ which is responsible for the symmetry breaking. However, the further flipping doesn't really lead to any problems, so it is not necessary to have such a principle in the first place.
\bibitem{f11}
Whether the scale happens to be above or below the QCD scale is irrelevant here since there are electrically neutral 
colored particles. This is unlike the minimal SM where even though the electric charge is quantized in 
multiples of 1/3 at distance scales smaller than the QCD scale, only integer charges are present at larger scales.
\bibitem{f12}
There are twice as many such fermions as the left-handed neutrino and the left-handed antineutrino is a linear combination of them.
\end{thebibliography}
\end{document}